\documentclass[fleqn,usenatbib]{mnras}

\usepackage{newtxtext,newtxmath}

\usepackage[T1]{fontenc}

\DeclareRobustCommand{\VAN}[3]{#2}
\let\VANthebibliography\thebibliography
\def\thebibliography{\DeclareRobustCommand{\VAN}[3]{##3}\VANthebibliography}

\def\orcid#1{\href{https://orcid.org/#1}{\includegraphics[scale=0.3]{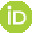}}}

\usepackage{graphicx}
\usepackage{amsmath}

\title[Six Massive Protostructures at $2.5<z<4.5$]{Identification and Characterization of Six Spectroscopically Confirmed Massive Protostructures at $2.5<z<4.5$}

\author[E. A. Shah et al.]{Ekta A. Shah\orcid{0000-0001-7811-9042},$^{1}$\thanks{E-mail: eashah@ucdavis.edu}
Brian Lemaux\orcid{0000-0002-1428-7036},$^{2,1}$
Benjamin Forrest\orcid{0000-0001-6003-0541},$^{1}$
Olga Cucciati\orcid{0000-0002-9336-7551},$^3$
Denise Hung\orcid{0000-0001-7523-140X},$^{2,4}$
Priti Staab\orcid{0000-0002-8877-4320},$^{1}$
\newauthor
Nimish Hathi\orcid{0000-0001-6145-5090},$^{5}$
Lori Lubin\orcid{0000-0003-2119-8151},$^{1}$
Roy R. Gal\orcid{0000-0001-8255-6560},$^{4}$
Lu Shen\orcid{0000-0001-9495-7759}$^{6,7}$
Giovanni Zamorani\orcid{0000-0002-2318-301X},$^{3}$
Finn Giddings\orcid{0009-0003-2158-1246},$^{4}$
\newauthor
Sandro Bardelli\orcid{0000-0002-8900-0298},$^{3}$
Letizia Pasqua Cassara\orcid{0000-0001-5760-089X},$^{8}$
Paolo Cassata\orcid{0000-0002-6716-4400},$^{9}$
Thierry Contini\orcid{0000-0003-0275-938X},$^{10}$
Emmet Golden-Marx\orcid{0000-0001-5160-6713},$^{11}$
\newauthor
Lucia Guaita\orcid{0000-0002-4902-0075},$^{12}$
Gayathri Gururajan\orcid{0000-0002-7472-7697},$^{13,3}$
Anton M. Koekemoer\orcid{0000-0002-6610-2048},$^{5}$
Derek McLeod\orcid{0000-0003-4368-3326},$^{14}$
Lidia A. M. Tasca\orcid{0000-0001-8560-5073},$^{15}$
\newauthor
Laurence Tresse\orcid{0000-0001-8776-0958},$^{16}$
Daniela Vergani\orcid{0000-0003-0898-2216},$^{3}$
Elena Zucca\orcid{0000-0002-5845-8132}$^{3}$
\\
$^{1}$ Department of Physics and Astronomy, University of California, Davis, One Shields Avenue, Davis, CA, 95616, USA\\
$^{2}$Gemini Observatory, 670 N. A’ohoku Place, Hilo, Hawai\textquotesingle i, 96720, USA\\
$^{3}$INAF-Osservatorio di Astrofisica e Scienza dello Spazio, Via Gobetti
93/3, I-40129, Bologna, Italy\\
$^{4}$University of Hawai\textquotesingle i, Institute for Astronomy, 2680 Woodlawn Drive, Honolulu, HI 96822, USA\\
$^{5}$Space Telescope Science Institute, Baltimore, MD 21218, USA\\
$^{6}$Department of Physics and Astronomy, Texas A\&M University, College Station, TX 77843-4242 USA\\
$^{7}$George P. and Cynthia Woods Mitchell Institute for Fundamental Physics and Astronomy, Texas A\&M University, College Station, TX 77843-4242 USA\\
$^{8}$INAF-IASF Milano, Via Alfonso Corti 12, 20159 Milano, Italy\\
$^{9}$Dipartimento di Fisica e Astronomia Galileo Galilei, Università degli Studi di Padova, Vicolo dell’Osservatorio 3, 35122 Padova Italy\\
$^{10}$Institut de Recherche en Astrophysique et Planétologie (IRAP), Université de Toulouse, CNRS, UPS, CNES, Toulouse, France\\
$^{11}$Department of Astronomy, Tsinghua University, Beijing 100084, China\\
$^{12}$Instituto de Astrofisica, Departamento de Ciencias Fisicas, Facultad de Ciencias Exactas,\\ Universidad Andres Bello, Fernandez Concha 700, Las Condes, Santiago RM, Chile\\
$^{13}$University of Bologna - Department of Physics and Astronomy “Augusto Righi” (DIFA), Via Gobetti 93/2, I-40129, Bologna, Italy\\
$^{14}$Institute for Astronomy, University of Edinburgh, Royal Observatory, Edinburgh, EH9 3HJ, UK\\
$^{15}$Laboratoire d’Astrophysique de Matseille\\
$^{16}$Aix-Marseille Univ., CNRS, CNES, LAM, 13388, Marseille Cedex 13, France\\
}

\date{Received October 12, 2023}

\pubyear{2023}

\begin{document}
\label{firstpage}
\pagerange{\pageref{firstpage}--\pageref{lastpage}}
\maketitle

\begin{abstract}

We present six spectroscopically confirmed massive protostructures, spanning a redshift range of $2.5<z<4.5$ in the Extended Chandra Deep Field South (ECDFS) field discovered as part of the Charting Cluster Construction in VUDS and ORELSE (C3VO) survey. We identify and characterize these remarkable systems by applying an overdensity measurement technique on an extensive data compilation of public and proprietary spectroscopic and photometric observations in this highly studied extragalactic field. Each of these six protostructures, i.e., a large scale overdensity (volume $>9000$\thinspace cMpc$^3$) of more than $2.5\sigma_{\delta}$ above the field density levels at these redshifts, have a total mass $M_{tot}\ge10^{14.8}M_\odot$ and one or more highly overdense (overdensity$\thinspace>5\sigma_{\delta}$) peaks. One of the most complex protostructures discovered is a massive ($M_{tot}=10^{15.1}M_\odot$) system at $z\sim3.47$ that contains six peaks and 55 spectroscopic members. We also discover protostructures at $z\sim3.30$ and $z\sim3.70$ that appear to at least partially overlap on sky with the protostructure at $z\sim3.47$, suggesting a possible connection. We additionally report on the discovery of three massive protostructures at $z=2.67$, 2.80, and 4.14 and discuss their properties. Finally, we discuss the relationship between star formation rate and environment in the richest of these protostructures, finding an enhancement of star formation activity in the densest regions. The diversity of the protostructures reported here provide an opportunity to study the complex effects of dense environments on galaxy evolution over a large redshift range in the early universe.

\end{abstract}




\section{Introduction}

Galaxy clusters are the most massive gravitationally bound systems in our universe. The processes driving their formation and their effect on the constituent galaxies, especially in the early universe, remain areas of ongoing research. To understand these processes and constrain their significance across cosmic time, studies of large populations of the progenitors of the massive clusters observed in the local universe are required. These progenitors are known as galaxy protoclusters\footnote{We use the more agnostic term  ``protostructures'' throughout the paper as we are unsure of the fate of the systems reported in this paper.}. Protoclusters are considered to be in the process of becoming gravitationally bound systems, finally collapsing into galaxy clusters by $z=0$ (or earlier). However, observational limitations constrain our ability to confirm if a given high-redshift protocluster candidate will eventually evolve into a present-day galaxy cluster. Therefore, many observationally-based studies use the definition of a protocluster as a structure with high-enough overdensity of galaxies (with respect to its surroundings) on large ($\sim10$\thinspace comoving Mpc) scales \citep{overzier2016}.

Studies have shown that dense environments play a critical role in galaxy evolution. At lower redshifts ($z<2$), through processes such as ram pressure stripping \citep{abadi1999,kenji2009,bosseli2022}, harassment \citep{moore1996,moore1998}, strangulation \citep{larson1980,bekki2002,vandenbosch2008}, viscous stripping \citep{nulsen1982}, and thermal evaporation \citep{cowie1977}, overdense environments in galaxy clusters accelerate galaxy evolution, making galaxies redder, and reducing or quenching their star formation compared to their counterparts in sparser (i.e., field) environments \citep[e.g.,][]{lemaux2019,tomczak2019,old2020,vanderburg2020,mcnab2021}. On average, there is over-representation of highly massive galaxies in clusters at $z$$\sim$1 compared to the field  \citep{baldry2006,bamford2009,calvi2013, tomczak2017}.

Given the result that massive galaxies with very low star formation rates (SFRs) are overrepresented in clusters at these redshifts, the implication is that the progenitors of such galaxies must have experienced rapid growth in the past to achieve their high stellar mass. This rapid growth is suggested by some studies showing higher SFRs in overdense protocluster galaxies compared to field galaxies at high redshift ($z>2$) \citep[e.g.,][though, see also \citealt{chartab2020}]{greenslade2018,miller2018,ito2020, lemaux2022,toshikawa2023}. The roles of various processes that can facilitate this rapid growth — such as mergers and interactions \citep{alonso2012,mei2023}, gas accretion \citep{damato2020}, interactions with the intracluster medium (ICM) \citep{dimascolo2023}, the contrast between in-situ and ex-situ stellar mass assembly \citep{cannarozzo2023}, star formation efficiency \citep{zavala2019,bassini2020}, active galactic nuclei (AGN) feedback \citep{brienza2023}, and AGN-ram pressure stripping connection \citep{peluso2022} - are yet to be fully understood. In order to unravel the complex interplay of processes guiding galaxy evolution within high-density environments and to discern how these processes evolve across cosmic time, large samples of high-redshift protostructures are needed.

While clusters of galaxies can be identified using various methods, finding protostructures can be more challenging. Many studies utilize relatively rare tracers, such as radio galaxies \citep{hatch2014,karouzos2014}, quasars \citep{song2016}, dusty star-forming galaxies (DSFGs) \citep{clements2014,casey2015,hung2016}, strong Lyman-alpha emitters (LAEs) \citep{shi2019,jiang2018} and ultra-massive galaxies \citep{mcConachie2022} to trace protostructures. However, some studies show no significant association between these tracers and protostructures \citep{husband2013,uchiyama2018}, and it is unclear that such tracers do not select a biased protocluster sample when they are found to be associated with an overdensity. Preferably, one would instead select samples of protostructures traced by galaxies that are representative of the overall population at a given epoch.

In this study, we leverage the plethora of observations in the Extended Chandra Deep Field South (ECDFS) field. This widely-studied extragalactic field contains extensive imaging \citep[e.g.,][]{wuyts2008, cardamone2010, dahlen2013,hsu2014} and spectroscopic data \citep[e.g.,][]{lefevre2004,lefevre2013,kriek2015,mclure2018}. These exquisite data, along with new spectroscopic observations taken as part of the Charting Cluster Construction in VUDS and ORELSE (C3VO; \citealt{lemaux2022}) survey, in concert with a novel density mapping technique allowed us to identify a large number of protostructures in the ECDFS field over the redshift range $2<z<5$. This density mapping technique, known as Voronoi Monte Carlo (VMC) mapping, has already been used to discover and/or characterize other massive protostructures: Hyperion at $z=2.5$ \citep{cucciati2018}, PCl J1000+0200 at $z=2.9$ \citep{cucciati2014}, PCl J0227-0421 at $z=3.3$ \citep{lemaux2014,shen2021}, Elent\'ari at $z=3.3$ \citep{forrest2023}, and PCl J1001+0220 at $z=4.6$ \citep[][Staab et al., \emph{submitted}]{lemaux2018}.

In this study, we present six of the most formidable protostructures in the ECDFS field found in our search over the redshift range $2.5<z<4.5$. These protostructures, with their wide range of redshift, mass, morphology and complexity offer a great opportunity for advancing our understanding of galaxy evolution during the critical epoch of the early universe.

\begin{figure*}
\includegraphics[scale=0.39]{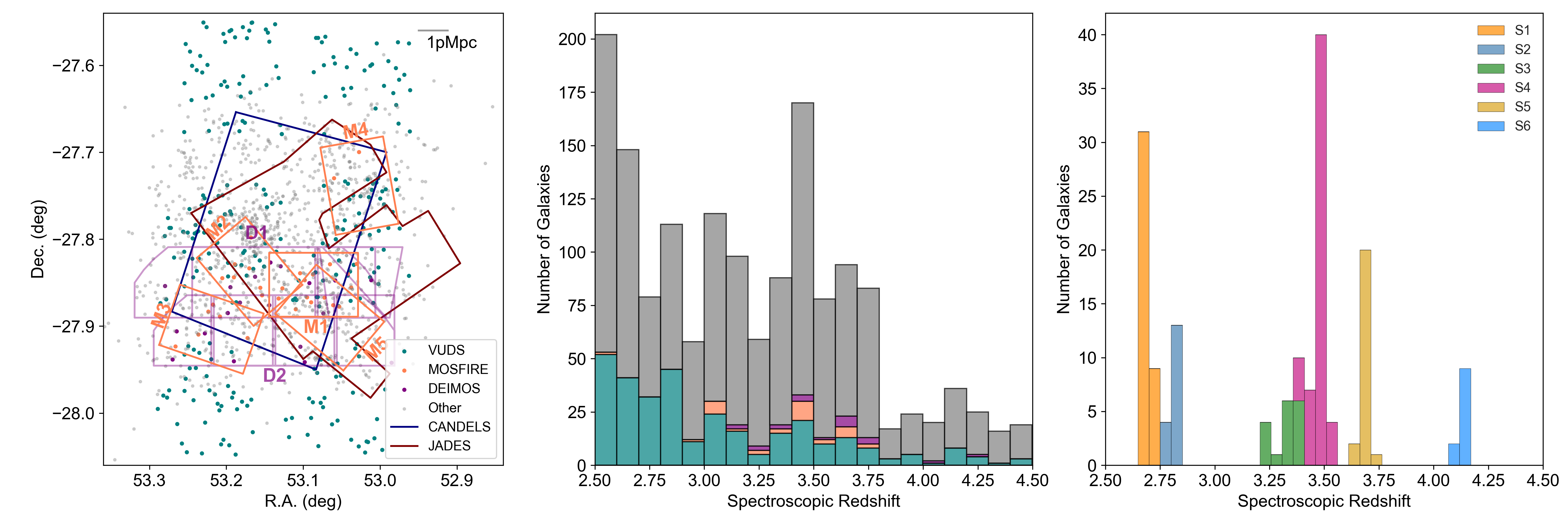}
    \caption{Left: The spatial distribution of galaxies with a secure spectroscopic redshift falling within the range of $2.5<z<4.5$. Galaxies are colored based on the survey from which the spectroscopic redshift was obtained: the VUDS survey (green/blue), C3VO MOSFIRE (orange) and DEIMOS (purple), and a compilation of other spectral surveys (gray, see text). The footprints of the GOODS-S and JADES surveys are also shown. Center: Stacked histogram representing the spectroscopic redshift distribution of all galaxies with a secure $z_{spec}$ in this range adopting the same color coding as in the left panel. Right: The redshift distribution of the spectroscopic members of each of the six different protostructures presented in this study as labeled in Figure ~\ref{fig:allstr2d}.}
    \label{fig:zdistr2z5}
\end{figure*}

The structure of this paper is as follows: spectroscopic and photometric data are described in \S\ref{sec:data}. In \S\ref{sec:char_of_str}, we discuss the methodology used to identify and characterize protostructures. In \S\ref{sec:ind_str}, we describe the individual protostructures along with their properties. We discuss our findings and compare them with other observational studies and expectations from simulations in \S\ref{sec:discussion}. Finally, in \S\ref{sec:summary}, we summarize our study. Throughout this study, we use a \citet{chabrier2003} IMF, AB magnitude system \citep{oke1983}, and a $\Lambda$CDM cosmology with $H_0=70$\thinspace km/s/Mpc, $\Omega_M=0.27$, and $\Omega_\Lambda=0.73$. Both comoving Mpc and proper Mpc distances are used in this study and are denoted cMpc and pMpc, respectively.

\section{Data}\label{sec:data}

The ECDFS \citep{lehmer2005} survey was envisioned as an expansion on the \textit{Chandra} Deep Field South survey \citep{giacconi2002} with 2\thinspace Ms of \textit{Chandra} X-ray observations \citep{virani2006,xue2016} across the entire field (and up to 7 Ms in some areas). It has now been targeted across the multi-wavelength spectrum \citep[e.g.,][]{zheng2004, grazian2006,wuyts2008, cardamone2010, luo2010,dahlen2013,hsu2014}, and become one of main targets for galaxy evolution studies (e.g., \citealt{kaviraj2008,lefevre2015,marchi2018,birkin2021}). This extended field spans an area of 0.5$^\circ\times\thinspace$0.5$^\circ$ in the southern sky. Here, we briefly describe the relevant photometric and spectroscopic data used in this work.

\subsection{Photometry}

In this study, we utilize the imaging and associated photometric catalogs from \citet{cardamone2010} and references therein. This catalog contains deep optical 18 medium-band photometry obtained using the Subaru telescope, combined with the existing UBVRIz obtained from the Garching-Bonn Deep Survey \citep[GaBoDS;][]{hildebrandt2006} and the Multiwavelength Survey by Yale-Chile  \citep[MUSYC;][]{gawiser2006} survey, deep near-infrared (NIR) imaging in JHK from MUSYC \citep{moy2003} and \textit{Spitzer} Infrared Array Camera (IRAC) images from the Spitzer IRAC/MUSYC Public Legacy Survey in ECDFS \citep[SIMPLE;][]{damen2011}. We selected the \citet{cardamone2010} catalog for our analysis after comparing it with an updated photometric catalog compiled by a deep VIMOS survey of the CDFS and UDS fields \citep[VANDELS;][]{mclure2018} team in which more contemporary observations were used. The VANDELS catalog consists of two catalogs in the CDFS field: VANDELS-HST and VANDELS-ground. These catalogs do not, however, cover the entirety of the VUDS footprint. A comparison of sources in the VANDELS catalogs and MUSYC \citep{cardamone2010} catalog over the area where the catalogs overlap shows that there is scatter at the faint end and this scatter seems to be different between the VANDELS-HST and VANDELS-ground catalog (i.e., there are lot of faint sources in the VANDELS-HST catalog) and this lack of homogeneity makes us weary of using a two catalog approach. For sources which are matched between the \citep{cardamone2010} and the VANDELS catalogs, the overall photometry, i.e., the apparent magnitudes in different bands and their associated errors, and the estimation of the physical parameters, such as, e.g., stellar mass and SFR, based performing SED fitting using the photometry from the various catalogs, were broadly comparable between the two catalogs. For example, the median offset in the stellar mass estimates using identical SED-fitting runs with \textsc{Le Phare} on the photometry from the \citep{cardamone2010} and VANDELS-ground catalog for galaxies with photometric redshift of $2.5 < z < 4.5$ was $\sim0.16$ dex. Despite the various virtues of the VANDELS photometric catalogs, such as having updated observations from \textit{HST} and VISTA, for our purposes we prioritized uniformity across the region we mean to reconstruct the density field. As such, we decided to retain the \citep{cardamone2010} catalog for this study. More details will be given in a companion paper (Shah et al.\ \emph{in prep}). Photometric redshifts ($z_{phot}$) were fit to the \citet{cardamone2010} photometry using the method described in \cite{{lefevre2015}} and references therein.

We estimate physical parameters of the galaxies, e.g., stellar mass and SFR, by using the spectral energy distribution (SED) fitting code \textit{LePhare} \citep{arnouts1999,ilbert2006} in conjunction with the \citet{cardamone2010} catalog, with the redshift of galaxies fixed to the $z_{phot}$ or $z_{spec}$ (when available, see next section). The adopted methodology is identical to that used in \citep{lefevre2015,tasca2015,lemaux2022}.

For this study, we only use photometric and spectroscopic objects with IRAC1 or IRAC2 magnitudes brighter than 24.8. This cutoff was selected based on the 3$\sigma$ limiting depth of the IRAC images in the ECDFS and a reliable detection in the rest-frame optical in order to constrain the Balmer/4000\AA\ break for galaxies at $2<z<5$. Adopting a similar method to \citet{lemaux2018}, we estimate the 80\% stellar mass completeness of our selected sample to be $log(M_*/M_\odot)\sim9.0-9.34$ (depending on the redshift). This stellar mass limit is additionally imposed on all $z_{spec}$ members reported in this paper.

\subsection{Spectroscopy}

Spectroscopic redshifts ($z_{spec}$) are crucial for mapping the underlying density field with a high degree of confidence. In this study, we employ a wide range of proprietary and publicly available spectroscopic observations in the ECDFS.

We use observations from Keck/DEep Imaging Multi-Object Spectrograph (DEIMOS, \citealt{faber2003}) and Keck/Multi-Object Spectrometer for Infra-Red Exploration (MOSFIRE, \citealt{mclean2010,mclean2012}) obtained as a part of the C3VO survey \citep{lemaux2022}. We targeted a suspected protostructure at $z\sim3.5$ \citep[e.g.,][]{ginolfi2017,forrest2017} using five MOSFIRE masks PClJ0332\_mask1-mask5 and two DEIMOS slitmasks: dongECN1 and dongECS1. Targeting for DEIMOS and MOSFIRE followed a similar prioritization scheme to that described in (\citealt{lemaux2022, forrest2023}, Staab et al. \emph{submitted}) and will be described in detail in our companion paper (Shah et al. \emph{in prep}).

For the DEIMOS observations, we used the GG400 order blocking filter with $\lambda_c=7000\mathring{A}$ and $1''$ wide slits. The total integration time was 4h45m and 2h10m for the masks dongECN1 and dongECS1, respectively, with an average seeing of $\sim0.9''$ and no extinction. The placements of these masks, labeled D1 and D2, respectively, are shown in the left panel of Figure~\ref{fig:zdistr2z5}. These data were reduced using a modified version of the spec2D pipeline \citep{cooper2012,newman2013} and analyzed using the technique described in \citet{lemaux2022}. For the MOSFIRE data, all observations were taken in the K band. The integration time ranges from 1h18m to 1h36m with a seeing range of $\sim0.65''-1.05''$ and little to no extinction for the five MOSFIRE masks, PClJ0332\_mask1-mask5. These masks are shown in the left panel of Figure~\ref{fig:zdistr2z5} and labeled M1-M5. These data were reduced using the MOSDEF2D data reduction pipeline \citep{kriek2015} and spectroscopic redshifts were measured adopting the method of \citet{forrest2023} and Forrest et. al. (\emph{in prep}). Additional details will be provided in a companion paper. In total, we recovered 29 and 26 secure (i.e., reliability of $\gtrsim95$\%) spectroscopic redshifts from the MOSFIRE and DEIMOS observations, respectively, with the vast majority of these redshifts in the range of $2.5<z<4.5$.

Other spectroscopic redshifts are incorporated from the the VIMOS Ultra-Deep Survey \citep[VUDS;][]{lefevre2015} and a list of publicly available redshifts compiled by one of the authors (NPH). The latter catalog contains spectroscopic redshifts from various surveys such as the VIsible Multi-Object Spectrograph \citep[VIMOS;][]{lefevre2003}- based the VIMOS VLT Deep Survey \citep[VVDS;][]{lefevre2004,lefevre2013}, the MOSFIRE Deep Evolution Field (MOSDEF) survey \citep{kriek2015}, the 3D-HST survey \citep{momcheva2016}, \citep[VANDELS;][]{mclure2018,pentericci2018}, and a variety of other surveys. These surveys usually target star-forming galaxies (SFGs) at $\sim$L$^*$ are broadly representative of SFGs at these redshifts with the exception of dusty galaxies (see discussion in \citeauthor{lemaux2022} \citeyear{lemaux2022}). In cases where we have more than one spectroscopic redshift for a given photometric object, we select the best $z_{spec}$ based on criteria such as redshift quality, instrument, survey depth, and photometric redshift (more details will be given in Shah et al. \emph{in prep}). After resolving duplicates, we retained 1539 unique galaxies with a secure $z_{spec}$ (in this case, corresponding to a reliability of $\gtrsim70$\%) over $2.5<z<4.5$, with 1075 of these galaxies satisfying the IRAC1/2 cut mentioned in the previous section.

Figure~\ref{fig:zdistr2z5}, shows the redshift and spatial distribution of all 1539 galaxies with a secure $z_{spec}$ in the range $2.5<z<4.5$. We also present the redshift distribution of all the $z_{spec}$ members of the protostructures reported in this work (described in the next two sections). In the left panel of Figure~\ref{fig:zdistr2z5}, we also show the footprints of the GOODS-S portion of the Cosmic Assembly Near-infrared Deep Extragalactic Legacy Survey  \citep[CANDELS;][]{grogin2011,koekemoer2011,guo2013} and the Near-Infrared Spectrograph (NIRSpec)- based observations taken as a part of the JWST Advanced Deep Extragalactic Survey \citep[JADES;][]{eisenstein2023}. These dedicated observations overlap with portions of the protostructures reported here, can be leveraged for more in-depth investigations in the future.

\section{Characterization of protostructures} \label{sec:char_of_str}

\subsection{Environment measurement using VMC-mapping}

We use Voronoi tessellation Monte Carlo (VMC) mapping to quantify the environment of galaxies. The VMC method is described in detail in a variety of other papers (e.g., \citealt{lemaux2017,lemaux2018, tomczak2017,cucciati2018,hung2020,shen2021}. The VMC mapping method divides the distribution of galaxies in cells called Voronoi cells based on their proximity with other galaxies. Hence it encapsulates the variation in galaxy distribution, making it a reliable measure of the local density of galaxies. We use both spectroscopic and photometric redshifts weighted based on their uncertainty to select redshifts for different Monte Carlo iterations. The exact version of VMC mapping used for this study is that of \citet{lemaux2022,forrest2023}.

The output of the VMC process is a measure of galaxy overdensity ($\delta_{gal}$) and the significance of overdensity ($\sigma_{\delta}$) for individual VMC cells over a 3D-grid along the RA-DEC and z (redshift) axis. For more details on how the latter is calculated, see \citet{forrest2023} and Staab et. al (\emph{submitted}). Overdensity values for galaxies are defined as the $\sigma_\delta$ value of the VMC cell that is closest to the galaxy coordinates.

\subsection{Defining and Identifying Large Protostructures Encapsulating Overdense Peaks}

We use the method described in \citet{cucciati2018,shen2021,forrest2023} to identify overdense peaks and their corresponding protostructures. These peaks and protostructures are defined as overdensity isopleths consisting of contiguous voxels with overdensity significance of $\sigma_{\delta}>5$ and $\sigma_{\delta}>2.5$, respectively. The coordinates and redshift of a given protostructure are defined as the the overdensity-weighted barycenter in each dimension of all contiguous voxels at $\sigma_{\delta}>2.5$ of a given protostructure (see more details later in this section.) Spectral members of a given protostructure are defined as those galaxies bounded by the $\sigma_{\delta}>2.5$ isopleths of that protostructure. The redshift bounds of the volume defined by the set of contiguous voxels that satisfy $\sigma_{\delta}>2.5$ for a given protostructure set the redshift bounds of that protostructure. 

In this paper, we present the six most massive (M$_{tot}\ge10^{14.8}M_\odot$)
protostructures in the $2.5<z<4.5$ identified in our sample using this method. All of the reported protostructures also get detected if we vary threshold from $2.5\sigma_{\delta}$ to $2\sigma_{\delta}-3\sigma_{\delta}$ (though the extension of the protostructures change), suggesting the detection of these protostructures is robust against changes in $\sigma_{\delta}$. In a companion paper, we will report on the full ensemble of the protostructures identified in this field.

Table~\ref{tab:allvals} reports the properties of these six protostructures and their corresponding peaks. The total mass of the protostructure (or peak) is calculated using $M_{tot}=\rho_{m}V(1+\delta_{m})$, where $\rho_{m}$ is the comoving matter density, $\delta_{m}$ is the mass overdensity, and V is the volume of the $2.5\sigma_{\delta}$ (or $5\sigma_{\delta}$) envelope, computed by adding together the volume of all the voxels in the envelope. We determine the mass overdensity ($\delta_{m}$) by scaling the average galaxy overdensity in the envelope, ($\delta_{gal}$) using a bias factor, i.e., $\delta_{m}=\delta_{gal}/bias$.
For this study, we derive the bias values from a linear interpolation of the numbers presented in \citet{chiang2013}, with the interpolation using the redshift of a given protostructure and stellar mass 80\% completeness limit of our sample at a given redshift. The latter is estimated following the methods described in Appendix B of \citet{lemaux2018}. The bias values used for individual protostructures are provided in the footnote of Table~\ref{tab:allvals}. Adopting bias factors from other works leads to a negligible change in the reported results. For the vast majority of cases in our protostructure sample, changing the $\sigma_{\delta}$ values by $10\%$ compared to the fiducial value of 2.5 used in this study, the mass estimates of the protostructures change by less than 0.1\thinspace dex, which is much less than 0.25\thinspace dex systematic uncertainty estimated based on the comparison between VMC-based mass estimates and true masses of structures in simulations  (Hung et al., \emph{in prep}). Note that in this study, we only report on peaks more massive than $M_{tot}>10^{12}M_{\odot}$.

We apply a method identical to previous C3VO works such as \citet{cucciati2018}, \citet{shen2021} and \citet{forrest2023}, to determine the barycenter positions of the peaks and protostructures and the elongation corrections for the peaks. To calculate the position of the barycenters, we use $X_{bc} = \Sigma_{i}(\delta_{gal,X_{i}}X_{i})/\Sigma_{i}(\delta_{gal,X_{i}})$ for $X = R.A.,Dec.,z$ and effective radius $R_{X}=\sqrt{\Sigma_{i}(\delta_{gal,X_{i}}(X_{i}-X_{bc})^2)/\Sigma_{i}(\delta_{gal,X_{i}})}$. The estimated effective radius in the $z$ (redshift) dimension ($R_{z}$) is usually elongated compared to that in the transverse dimensions as they get affected by the relatively large uncertainties in the photometric redshifts as well as the peculiar velocities of galaxies in protostructures. Due to these effects, the measured value of $R_{z}$ is inflated compared to its intrinsic value. To correct for this effect on the volume and density estimation, we use an elongation correction factor $E_{z/xy} = R_{z}/R_{xy}$, where $R_{xy} = (R_x+R_y)/2$. The intrinsic (corrected) volume of the peak is then calculated as the ratio of the measured volume to the elongation factor $(V_{corr}=V_{meas}/E_{z/xy}).$ We also apply this correction to estimate the elongation corrected average overdensity using $<\delta_{gal}>_{corr}=M_{tot}/(V_{corr\rho_m})-1$. We only make these elongation-based corrections in these estimates of the properties of the peaks.

\begin{figure*}
\includegraphics[scale=0.87]{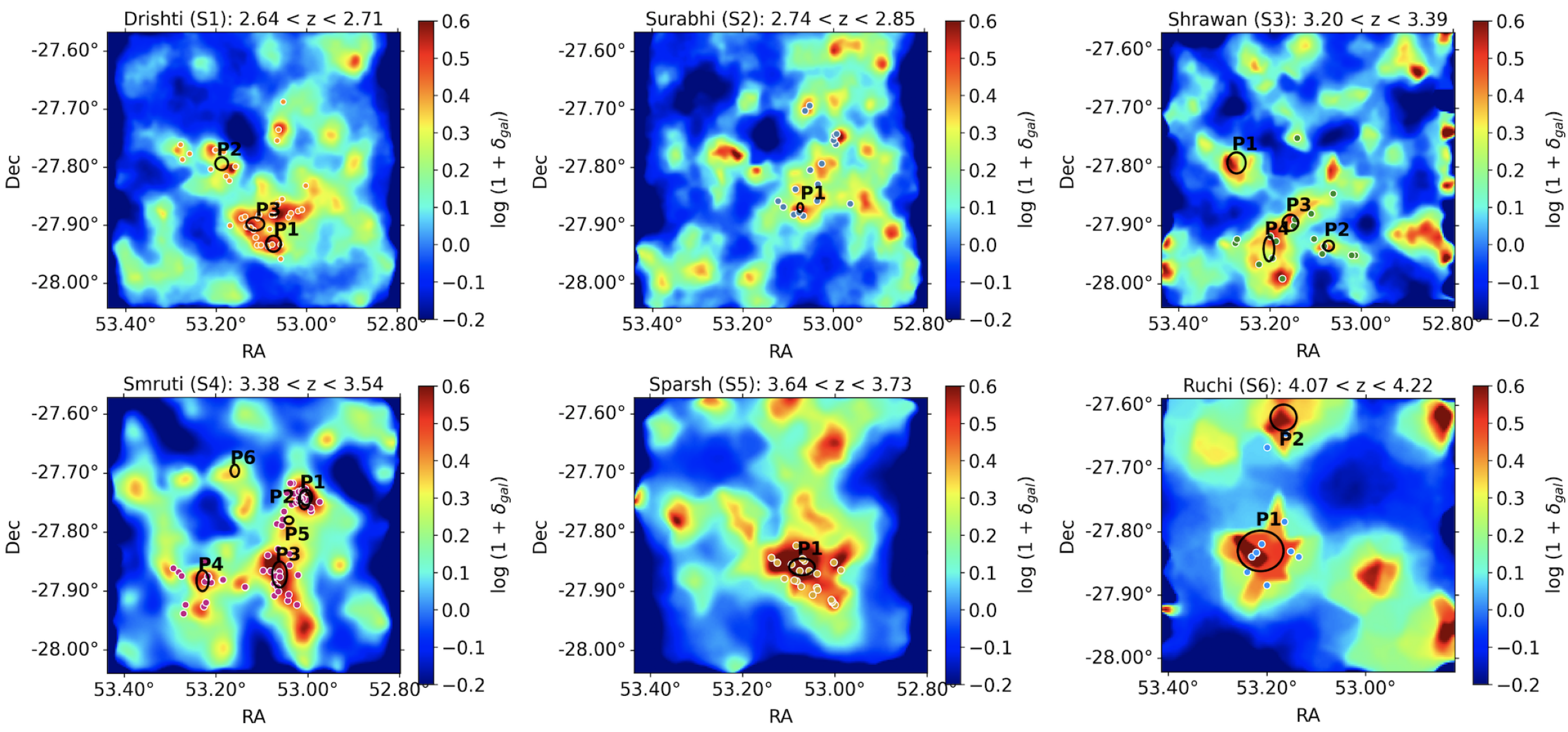}
    \caption{Projected Overdensity map of all six protostructures ($\sigma_{\delta}>2.5$) at $2.5<z<4.5$ presented in this study: The darker red colors present higher overdensity values and bluer colors present  lower overdensity values. The overdense peaks ($\sigma_{\delta}>5$) with mass log($M/M_\odot>12$) are represented with black ellipses based on their $R_x$ and $R_y$ values from Table~\ref{tab:allvals}. The spectroscopic members of protostructures are presented using dots with the same protostructure specific color scheme as in the third panel of Figure~\ref{fig:zdistr2z5}.}

    \label{fig:allstr2d}
\end{figure*}

\begin{figure*}
\begin{center}
\includegraphics[scale=0.494, clip=true]{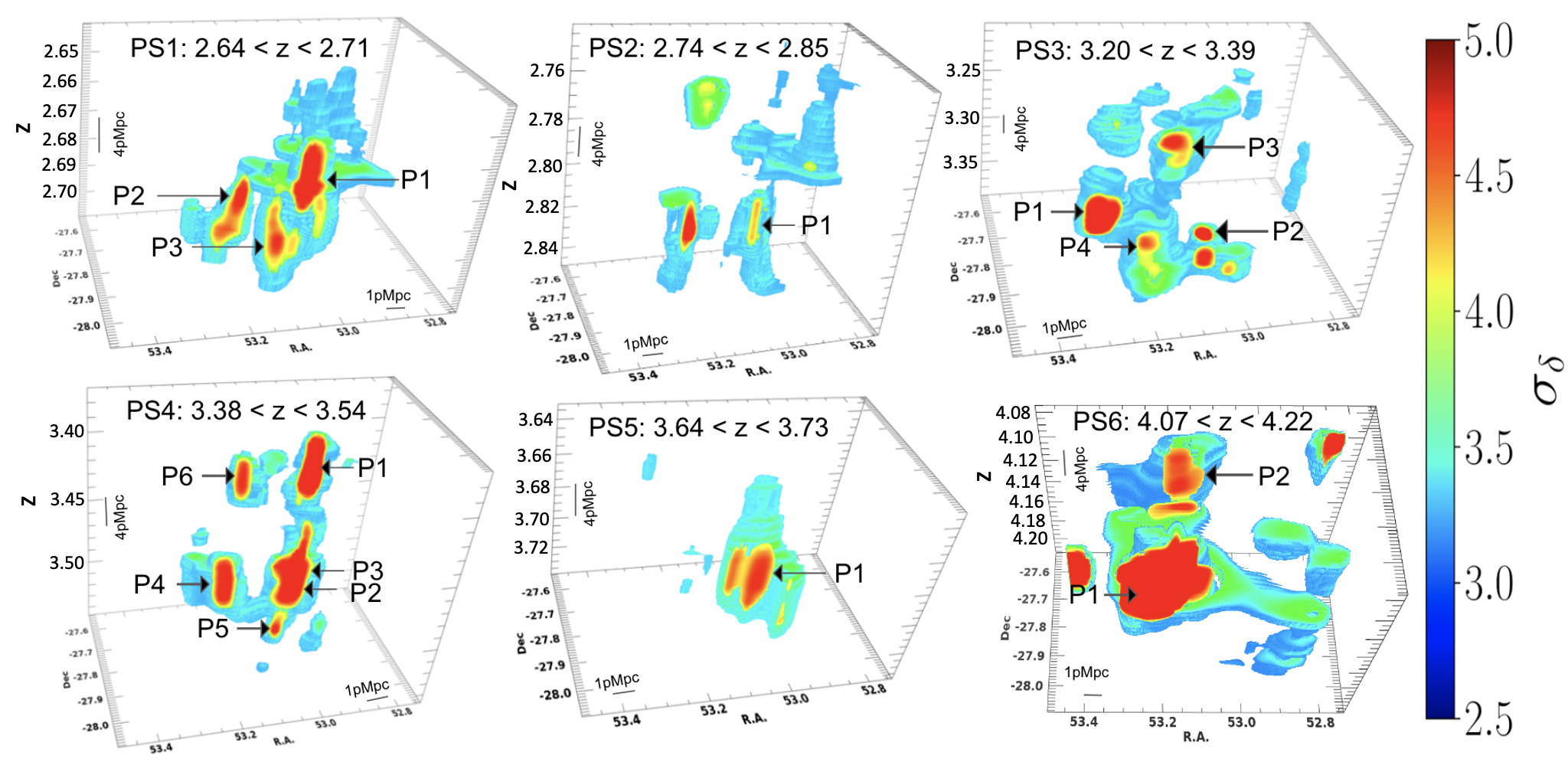}
    \caption{3D Overdensity map of ECDFS in the redshift range  all six protostructures: Red color shows higher overdensity and light blue shows lower overdensity. As a reminder, we use the term protostructure here and throughout the paper agnostically as we are unsure of their future fate. It can be seen that in some protostructures (e.g., protostructure 4 in lower left and protostructure 2 in the middle column of the upper panel), the overdense regions in red are connected through a relatively lower density bridge. There is also a wide range in the morphology and volume of the protostructures. The range of $\sigma_{\delta}$ is slightly larger (2.5-6.0\thinspace$\sigma_{\delta}$) for the highest redshift protostructure (PS6) than shown in the colorbar for a better visual representation of the structure.}
    \label{fig:allstr3d}
\end{center}
\end{figure*}

We report associated quantities for all six protostructures detailed in this work in Table~\ref{tab:allvals}. The 2D and 3D overdensity maps of the six protostructures are presented in Figure~\ref{fig:allstr2d} and Figure~\ref{fig:allstr3d}, respectively. We also show the redshift distribution of the $z_{spec}$ members of the protostructures in the right panel of Figure~\ref{fig:zdistr2z5}. We describe these six protostructures and their properties below.

\begin{table*}
\caption{The properties of all six protostructures (S1-S6) and their corresponding overdense peaks (Pi) estimated using formulae described in Section~\ref{sec:char_of_str}.}
\begin{tabular}{lllllllllllllll}

\hline
ID     & RA      & DEC      & z &$n_{sp}$$^{a}$& \textless{}$\delta_{gal}$\textgreater{}$^{b}$& V       & $logM_{tot}$& SzF$^{c}$    & Rx$^{d}$     & Ry$^{d}$ & Rz$^{d}$ & Ez/xy$^{e}$ & V$_{corr}^{f}$  & \textless{}$\delta_{gal}$\textgreater{}$_{corr}$$^{f}$\\
\hline
S1     & 53.0824 & -27.8670 & 2.671 & 40 & 1.21 & 11292 & 14.9 & 0.09 & - & -  & -  & -    & -  & -                                 \\
P1\_S1 & 53.0731 & -27.9323 & 2.674  & -& 3.03                         & 495    & 13.7   & -      & 1.51   & 1.27   & 6.12   & 4.40  & 127  & 20.30                              \\
P2\_S1 & 53.1876 & -27.7943 & 2.694  & -& 2.36                         & 297    & 13.4   & -      & 1.49   & 1.16   & 7.54   & 5.69  & 59   & 23.11                            \\
P3\_S1 & 53.1133 & -27.8984 & 2.697  & -& 2.12                         & 381    & 13.5   & -      & 2.09   & 1.19   & 6.58   & 4.01  & 107  & 14.97                              \\
\hline
S2     & 52.9988 & -27.8063 & 2.795  & 17 & 0.95                         & 11251 & 14.8   & 0.09 & -      & -      & -      & -     & -       & -                                 \\
P1\_S2 & 53.0731 & -27.8694 & 2.809  & -& 1.97                         & 111   & 12.9   & -      & 0.61   & 0.70   & 6.60   & 10.09 & 12   & 39.05                             \\
          \hline
S3     & 53.1519 & -27.9222 & 3.301  & 17 & 0.90                         & 23634 & 15.1   & 0.12  & -      & -      & -      & -     & -       & -                                 \\
P1\_S3 & 53.2727 & -27.7936 & 3.343  & -& 2.47                         & 1683  & 14.1   & -      & 2.05   & 1.85   & 8.53   & 4.36  & 386  & 19.01  
           \\
P2\_S3 & 53.0714 & -27.9353 & 3.355  & -& 1.94                         & 263   & 13.3   & -      & 1.17   & 0.91   & 4.98   & 4.80  & 55   & 18.62      
           \\
P3\_S3 & 53.1552 & -27.8959 & 3.242  & -& 1.98                         & 629   & 13.7   & -      & 1.73   & 1.38   & 7.06   & 4.54  & 139  & 17.66  
            \\
P4\_S3 & 53.2022 & -27.9406 & 3.335  & -& 1.63                         & 483   & 13.5   & -      & 1.19   & 2.16   & 7.81   & 4.66  & 103  & 16.59    
           \\
\hline
S4     & 53.0848 & -27.8250 & 3.466  & 55 &  1.75                         & 19854 & 15.1   & 0.18  & -      & -      & -      & -     & -       & -                                 \\
P1\_S4 & 53.0076 & -27.7463 & 3.410  &  -& 3.18                         & 867   & 13.9   & -      & 1.23   & 1.60   & 9.67   & 6.83  & 127  & 36.62     
           \\
P2\_S4 & 53.0042 & -27.7411 & 3.479  &-& 3.86                         & 650   & 13.8   & -      & 1.34   & 1.37   & 10.03  & 7.40  & 88  & 44.90  
           \\
P3\_S4 & 53.0613 & -27.8723 & 3.471  & -& 3.70                         & 1740  & 14.3   & -      & 1.68   & 2.30   & 9.44   & 4.75  & 367  & 27.11    
            \\
P4\_S4 & 53.2290 & -27.8828 & 3.462  & -& 3.08                         & 745   & 13.8   & -      & 1.37   & 1.84   & 8.95   & 5.57  & 134  & 28.82  
           \\
P5\_S4 & 53.0412 & -27.7804 & 3.530  & -& 3.93                         & 141   & 13.2   & -      & 0.99   & 0.66   & 5.25   & 6.36  & 22   & 38.66          
           \\
P6\_S4 & 53.1586 & -27.6964 & 3.418  & -& 2.32                         & 268   & 13.3   & -      & 0.95   & 1.06   & 6.05   & 6.04  & 44   & 26.93                            
            \\ \hline
S5     & 53.0579 & -27.8670 & 3.696  & 22 & 2.36                         & 9032 & 14.8   & 0.20  & -      & -      & -      & -     & -       & -                                 \\
P1\_S5 & 53.0714 & -27.8592 & 3.696  & -& 4.46                         & 1201  & 14.1   & -      & 2.96   & 1.52   & 7.17   & 3.20  & 375  & 20.22                              \\
\hline
S6     & 53.1876 & -27.7991 & 4.144 & 11 & 1.15                         & 42319 & 15.4   & 0.14 & -      & -      & -      & -     & -       & -                                 \\
P1\_S6 & 53.2124 & -27.8306 & 4.150  & -& 2.48                         & 11748 & 15.0   & -      & 5.20   & 3.60   & 10.79  & 2.45  & 4789 & 10.51                              
            \\
P2\_S6 & 53.1659 & -27.6199 & 4.109  & -& 1.28                         & 2552  & 14.2   & -      & 2.96   & 2.31   & 8.99   & 3.41  & 748  & 11.68                              \\
\hline
\end{tabular}
\footnotesize
   The units of the columns of the table are - RA, DEC: [deg], V, Vcorr: [cMpc$^3$], $M_{tot}$: [$M_{\odot}$], and Rx, Ry, and Rz: [cMpc]. The bias values used for the calculation of $M_{tot}$, $V_{corr}$, and  \textless{}$\delta_{gal}$\textgreater{}$_{corr}$ for the six protostructures, in order of increasing protostructure redshift, are 2.05, 2.10, 2.45, 2.55, 2.70, and 3.02, respectively, and are based on  \citet{chiang2013}.
\begin{flushleft}
$a$: The number of $z_{spec}$ members in the protostructure satisfying stellar mass and IRAC magnitude cuts. We do not report the number of $z_{spec}$ members for peak regions
$b$: The average galaxy overdensity in the region of interest as measured on the VMC maps
$c$: Fraction of objects with photometric redshifts consistent with the protostructure that have secure spectroscopic redshifts
$d$: Effective radius of the region of interest in the transverse and line of sight dimensions
$e$: Elongation correction (see \citealt{cucciati2018})
$f$: Corrected for elongation
\end{flushleft}

\label{tab:allvals}
\end{table*}

\section{Individual Protostructures and their properties }\label{sec:ind_str}

\subsection{Protostructure 1: \textit{Drishti}}

\textit{Drishti}\footnote{We named the six protostructures after the 5+1 senses through which we perceive and experience the universe. These names are: \textit{Drishti} (vision), \textit{Surabhi} (fragrance), \textit{Shrawan} (hearing),
and \textit{Smruti} (intuition/memory) - collective wisdom transcending time and embedded in our DNA, \textit{Sparsh} (touch), and \textit{Ruchi} (taste, in Telugu). All names, except for \textit{Ruchi}, are in Sanskrit.} is the lowest redshift protostructure reported here. It is located at $[\alpha_{J2000}, \delta_{J2000}] = [53.0824, -27.8670$], spans $2.64<z<2.71$ and has a systemic redshift of $z$=2.671. It has a total mass of $10^{14.9}M_{\odot}$, an average $\sigma_\delta$ of 3.68, and occupies volume of 11292\thinspace cMpc$^3$.
It consists of three overdensity peaks, each with M$_{tot}>10^{13.3}M_{\odot}$ as shown in Figure~\ref{fig:allstr2d} and Figure~\ref{fig:allstr3d}. The southern-most peak P1\_S1 is the largest and most massive of the three peaks. This protostructure was suggested by \citet{guaita2020} based on the VANDELS observations. Their reported center of the highest density peak ($z=2.69$) is separated by $\sim3.4$$\arcmin$ ($\sim$1.6\thinspace pMpc in projection) from P3\_S1 at $z\sim2.697$. However, they did not have any $z_{spec}$ members for this protostructure as opposed to the 40 $z_{spec}$ members in this work.

\subsection{Protostructure 2: \textit{Surabhi}}
\textit{Surabhi} is located at $[\alpha_{J2000}, \delta_{J2000}] = [52.9988, -27.8063$] and $z$=2.795 ($2.74<z<2.85$). It has a total mass of $10^{14.8}M_{\odot}$, an average $\sigma_\delta$ of 3.29, and occupies a volume of $\sim11251$\thinspace cMpc$^3$. It has one overdensity peak with mass $>10^{12.8}M_{\odot}$, and two less massive peaks not reported here.

This protostructure may be related to three protoclusters at $z\sim2.8$ in ECDFS identified in \citet{zheng2016} based on the overdensity of LAEs. \citet{guaita2020} also report a protostructure at $z\sim 2.8$ that could be related to this protostructure. Their protostructure is located $\sim$5$\arcmin$ ($\sim$2.4\thinspace pMpc in projection) from P1\_S2 at $z\sim2.809$. They report four $z_{spec}$ members as compared to 17 spectroscopic members in our work.

\subsection{Protostructure 3: \textit{Shrawan}}
 \textit{Shrawan} is a massive protostructure situated at $[\alpha_{J2000}, \delta_{J2000}] = [53.1519, -27.9222$] and redshift $z$=3.301 ($3.20<z<3.39$). It has a total mass of $10^{15.1}M_{\odot}$, an average $\sigma_\delta$ of 3.45, and it encompasses 23634\thinspace cMpc$^3$. It contains four massive overdense peaks (each with $M_{tot}>10^{13.2}M_{\odot}$) as shown in Fig~\ref{fig:allstr2d} and Fig~\ref{fig:allstr3d}. The northern-most peak, P1\_S3, is the largest and most massive peak out of all four peaks. This protostructure has 17 spectroscopic member galaxies. A candidate overdensity, `CCPC-z32–003' at $z=3.258$, is reported in \citet{franck2016} at a similar location, though with a ``cluster probability'' of 10\%. This candidate is $\sim0.13$ deg ($\sim$3.5\thinspace pMpc projected) from the nearest peak, P3\_S3, at $z\sim3.24$. Another candidate, `CCPC-z33–003', is reported at $z=3.368$ is $\sim0.16$\thinspace deg ($\sim$4.3\thinspace pMpc in projection) from the nearest redshift overdensity peak P2\_S3 at $z\sim3.355$. However, this candidate has a similarly low cluster probability of 10\%.
 
\subsection{Protostructure 4: \textit{Smruti}}
\textit{Smruti} is a massive protostructure located at $[\alpha_{J2000}, \delta_{J2000}] = [53.0848, -27.8250$] and $z$=3.466 ($3.38<z<3.54$). It has a mass of $M_{tot}=10^{15.1}M_{\odot}$, an average $\sigma_\delta$ of 4.05, and occupies a volume of 19854\thinspace cMpc$^3$. It has six massive overdensity peaks (each with $M_{tot}>10^{13.1}M_{\odot}$) as shown in Fig~\ref{fig:allstr2d} and Fig~\ref{fig:allstr3d}. The protostructure contains 55 spectroscopic member galaxies.

This existence of this protostructure was suggested by a few studies. An overdensity of galaxies at $z\sim3.5$ was observed in the full redshift (both photometric and spectroscopic) distribution of galaxies in the GOODS-S/CDFS field in 3D-HST \citep{skelton2014}, as well as observations from The FourStar Galaxy Evolution Survey (ZFOURGE) \citep{straatman2016}. The overdensity was also alluded to in \citet{guaita2020} as a protocluster candidate at $z = 3.43$ identified in VANDELS with six spectroscopic members. It is $\sim$0.55$\arcmin$ ($\sim$0.24\thinspace pMpc in projection) away from the P1\_S4, suggesting they are part of the same protostructure. \citet{forrest2017} also detected an overdensity of Extreme [OIII]+H$\beta$ Emission Line Galaxies (EELGs) and Strong [OIII] Emission Line Galaxies (SELGs) at $z\sim3.5$ that is $\sim8.4$$\arcmin$ ($\sim$3.70\thinspace pMpc in projection) away from P3\_S4. \citet{franck2016} report the candidate `CCPC-z34–002' at $z=3.476$ with a cluster probability of 48\%, which is $\sim$0.9$\arcmin$ ($\sim$0.40\thinspace pMpc in projection) away from P3\_S4.  

The peak P3\_S4 also contains the most massive galaxy out of all ALMA-detected galaxies at $3<z<4$ in the GOODS-ALMA field \citep{ginolfi2017}. 
\citet{zhou2020} report that four optically dark galaxies detected in an ALMA continuum survey, reside in this protostructure, which suggests that considerable star formation activity is occurring in this protostructure. While the above studies appeared to detect parts of this protostructure, the extensive spectroscopic data and density mapping technique employed here interconnected and expanded on these detections.

\subsection{Protostructure 5: \textit{Sparsh}}
\textit{Sparsh} is located at $[\alpha_{J2000}, \delta_{J2000}] = [53.0579, -27.8670$] and $z=3.696$ ($3.64<z<3.73$). It has a mass of $10^{14.8}M_{\odot}$, an average $\sigma_\delta$ of 3.48, and occupies a volume of 9032\thinspace cMpc$^3$.  It contains one massive overdensity peak, as well as two lower mass ($<$10$^{13}$ $M_\odot$) peaks that are not reported here due to their small volume ($\sim$50 cMpc).

Hints of this protostructure were reported in \citet{kang2009}. They reported and overdensity at $z\sim3.7$ that is $\sim2.00$$\arcmin$ ($\sim$0.86\thinspace pMpc in projection) away from P1\_S5. This study was followed by \citet{kang2015}, who also report on the same candidate with two $z_{spec}$ members. We find 22 spectroscopic member galaxies in this protostructure. \citet{franck2016} have two candidates that may correspond to this protostructure. The first is `CCPC-z36–002' at $z=3.658$ with cluster probability 1\%, which is $\sim2.92\arcmin$ ($\sim1.26$\thinspace pMpc in projection) away from P1\_S5. The second is `CCPC-z37–001' at $z=3.704$ with cluster probability 10\%, which is $\sim10.75$$\arcmin$ ($\sim$4.6\thinspace pMpc in projection) away from P1\_S5.

\subsection{Protostructure 6: \textit{Ruchi}}
\textit{Ruchi} is the highest redshift protostructure reported here. It is located at $[\alpha_{J2000}, \delta_{J2000}] = [53.1876, -27.7991$] and $z\sim4.14$ ($4.07<z<4.22$). It is also the most massive protostructure, with a mass of $10^{15.4}M_{\odot}$, an average $\sigma_\delta$ of 5.15, and occupying the largest volume of our sample (42319\thinspace cMpc$^3$). It has two overdensity peaks, each with mass more than $10^{14.0}M_{\odot}$ as shown in Fig~\ref{fig:allstr2d} and Fig~\ref{fig:allstr3d}. We note that the precision of the mass estimates decreases at these high redshifts, due to relatively limited number of spectral redshifts and our inability to probe galaxies with lower luminosity and lower mass. There are 11 spectroscopic member galaxies in this protostructure. To our knowledge, this structure has not been reported in any other works.

\section{Discussion}\label{sec:discussion}

We report six massive protostructures (with masses greater than $10^{14.8}M_{\odot}$) in the ECDFS.  
While some hints of these structures were previously mentioned in other studies as described in the last section, it is only through our extensive spectroscopic and photometric samples, combined with the VMC mapping technique, that we have unequivocally confirmed the existence of these structures, mapped out their full extent, and measured their properties.

To contextualize these findings, we compare the observed number density of these protostructures with the predictions from a simulation based study that will be described in detail in an upcoming C3VO paper (Hung et al., \emph{in prep}). Very briefly, we employ the GAlaxy Evolution and Assembly (GAEA) semi-analytic (SAM) model \citep{xie2017} applied to the dark matter merger trees of Millenium Simulation \citep{springel2005}. A lightcone of radius 2.3\thinspace deg was generated from the Millenium simulation using a method similar to \citet{zoldan2017}. Mock observations are made of this lightcone that mimic the properties of the spectroscopic and photometric data in the ECDFS field, including spectroscopic redshift fraction and photo-$z$ statistics as a function of both redshift and apparent (IRAC1) magnitude. For each mock dataset, we perform identical VMC mapping to that performed on the real data (observations) and a structure finding technique is applied following the methodology described in Hung et al., \emph{in prep}. From the full lightcone, we sampled 1000 iterations of fields with a volume equivalent to that of ECDFS over the range 2.5$<$$z$$<$4.5 taking care to avoid boundary effects. For each sampled volume, we counted the number of simulated protoclusters with similar masses to those of the protostructures detected in ECDFS taking into account completeness effects. Over all 1000 iterations we recover an expectation value of five protostructures with similar masses to those reported in this paper within an ECDFS-like volume from 2.5$<$$z$$<$4.5, with a 1$\sigma$ range of 3-7. These numbers are well consistent with the number of massive protostructures recovered in our observations.

Additionally, the sizes of all protostructures, with the possible exception of the $z\sim4.144$ protostructure, are in agreement with the sizes predicted for protoclusters for the same mass and redshift based on simulations \citep[][, Hung et al., \emph{in prep}]{chiang2013,chiang2017, muldrew2015,contini2016}. For example, for simulated protoclusters in the mass and redshift range of the protostructures detailed in this work, \citet{chiang2013} report a range of effective radii of $\sim 4.5$ cMpc - $\sim$ 10.5 cMpc, which is a size scale comparable to the protostructures identified here. However, we caution the reader that many differences exist in the methods used for identification of structure in observations and simulations, as well as differences in how structure sizes are calculated between simulations and observations. Some of the nuances associated with these comparisons will be detailed in an upcoming work (Hung et al., \emph{in prep}). The range of volumes of all of our peaks at $z<4$ are comparable to the volume of the peaks of other C3VO structures at similar redshifts, e.g., Hyperion
\citep{cucciati2018}, Elent\'ari
\citep{forrest2023}, and PCl J0227-0421 \citep{shen2021}. The range of volumes for the highest redshift protostructure, \textit{Ruchi}, at $z\sim4.1$ is comparable to the range of volume of peaks in PCl J1001+0220 at $z\sim4.57$ (Staab et. al., \emph{submitted}).

\begin{figure}
\includegraphics[scale=0.56]{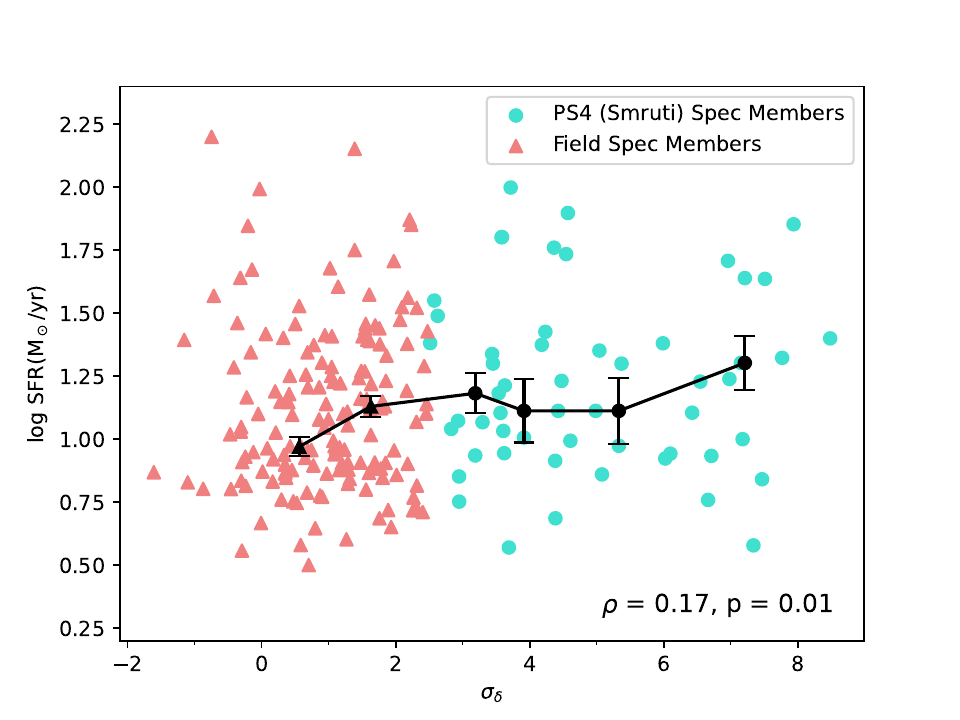}
    \caption{Relationship between SFR and overdensity ($\sigma_\delta$) for the spectroscopic members of the protostructure 4, i.e, \textit{Smruti} ($\sigma_\delta>2.5$ and $3.38<z<3.54$), presented using filled turquoise circles and the spectroscopic members of a corresponding coeval field sample at $3.2<z<3.7$ and $\sigma_\delta<2.5$ presented using coral triangles. The black points show median values of the SFRs in a given $\sigma_\delta$ bin. The error bars show $1\sigma$ uncertainties in the median SFR values. The $\sigma_\delta$ bins for the black points are created such that all bins in the coeval field have approximately the same number of points and all bins for the protostructure members also have approximately the same number of points. A Spearman test for this data show a weak but statistically significant positive correlation between the SFR and environment.}
    \label{fig:sfr_vs_odens_ps4}
\end{figure}
To understand the impact of the dense protostructure environments on galaxy evolution, as a test case, we focus on \textit{Smruti} at $z\sim3.5$, as elevated star formation in this protostructure has been hinted at in previous works \citep[e.g,][]{forrest2017,ginolfi2017,zhou2020}. SFRs of the 55 $z_{spec}$ members of the protostructure ($\sigma_\delta>2.5$) as well as a corresponding coeval field sample at $3.2<z<3.7$ and $\sigma_\delta<2.5$ were used to investigate the relationship between SFR and environment in this protostructure as shown in Figure~\ref{fig:sfr_vs_odens_ps4}. A Spearman test is performed and returns a correlation coefficient of $\rho=0.17$ with $p=0.01$, which implies a weak but statistically significant positive correlation. This correlation is $\sim$30\% stronger than that of the overall galaxy population at these redshifts \citep{lemaux2022}. The positive correlation indicates rapid \emph{in situ} stellar mass growth in the dense environments of such high-redshift protostructures, which is potentially necessary for forming the massive galaxies observed in clusters in the nearby universe \citep{baldry2006,bamford2009,calvi2013}. This enhanced star formation in protostructures at high redshift is also in agreement with the results from some simulation-based studies \citep[e.g.,][]{chiang2017} and observations-based studies \citep[e.g,][]{greenslade2018, lemaux2022}. We will present a detailed study on this relation, as well as other galaxy properties, in all of the protostructures reported here, as well as lower mass systems, in a follow-up paper (E. Shah et. al., \emph{in prep}).  

\section{Summary}\label{sec:summary}
We identify and present six spectroscopically-confirmed massive ($M_{tot}>10^{14.8}M_{\odot}$) protostructures at $2.5<z<4.5$ in the ECDFS field. These structures are identified by applying an overdensity-measurement technique on the publicly available extensive spectroscopic and photometric observations as well as targeted spectral observations in the ECDFS field from the C3VO survey. We calculate the volume, mass, and average overdensities of the protostructures, as well as other associated quantities. One of these protostructures, named \textit{Smruti}, is a large complex protostructure at $z\sim3.47$ containing six overdense ($\sigma_{\delta}>5$) peaks and 55 spectroscopic members. Its member galaxies show a statistically significant correlation between the SFR and environment density. This protostructure, as well as another protostructure at $z$$\sim$3.3, dubbed \textit{Shrawan}, are very massive ($M\sim10^{15.15}M_\odot$) and each contains $\ge4$ overdense peaks. The remaining protostructures at $z$$<$4 are slightly less massive (10$^{14.8-14.9}$ $M_{\odot}$) and contain fewer peaks. The highest redshift protostructure reported here ($z$$\sim$4.14), dubbed $\textit{Ruchi}$, contains 11 spectroscopic members. The number density, masses, and sizes of these protostructures are broadly in agreement with the prediction of these properties of protoclusters in simulations \citep[][, Hung et al., \emph{in prep}]{chiang2013,chiang2017,muldrew2015,contini2016}. These protostructures span wide ranges of complexity, masses, volume, and redshift and will be used in a companion paper (Shah et al. \emph{in prep}) to study the effect of dense environments on star formation and nuclear activity at high redshift.

\section{Data Availability}
The data used this study would be shared based on a reasonable request to the corresponding author.

\section*{Acknowledgements}

We are grateful to the anonymous reviewer for their insightful review of our manuscript. Their detailed comments have substantially enhanced the quality of the paper. Results in this paper were partially based on observations made at Cerro Tololo Inter-American Observatory at NSF’s NOIRLab, which is managed by the Association of Universities for Research in Astronomy (AURA) under a cooperative agreement with the National Science Foundation. Results additionally relied on observations collected at the European Organisation for Astronomical Research in the Southern Hemisphere. This work is based on observations made with the Spitzer Space Telescope, which is operated by the Jet Propulsion Laboratory, California Institute of Technology under a contract with NASA. Some of the data presented herein were obtained at Keck Observatory, which is a private 501(c)3 non-profit organization operated as a scientific partnership among the California Institute of Technology, the University of California, and the National Aeronautics and Space Administration. The Observatory was made possible by the generous financial support of the W. M. Keck Foundation. Some of the material presented in this paper is based upon work supported by the National Science Foundation under Grant No. 1908422. This work was additionally supported by NASA’s Astrophysics Data Analysis Program under grant number 80NSSC21K0986. GG acknowledges support  from the grants PRIN MIUR 2017 \texttt{\char`-}20173ML3WW\texttt{\char`_}001, ASI n.I/023/12/0 and INAF\texttt{\char`-}PRIN 1.05.01.85.08.
The authors wish to recognize and acknowledge the very significant cultural role and reverence that the summit of Maunakea has always had within the indigenous Hawaiian community. We are most fortunate to have the opportunity to conduct observations from this mountain. For the purpose of open access, the author has applied a Creative Commons Attribution (CC BY) licence to any Author Accepted Manuscript version arising from this submission.



\bibliographystyle{mnras}
\bibliography{6ps_ecdfs} 




\bsp
\label{lastpage}
\end{document}